\documentclass{emulateapj} 
 
\usepackage[dvipdf]{epsfig} 
\usepackage[dvips]{rotating}
\usepackage{subfigure}

\bibpunct{(}{)}{;}{a}{}{,} 
 
\shorttitle{AllWISE+NEOWISER Coadds}
 
\shortauthors{Meisner et al.} 
 
\begin{document} 
\title{Full-depth Coadds of the WISE and First-year NEOWISE-Reactivation Images}

\author{Aaron M. Meisner\altaffilmark{1,2}}
\author{Dustin Lang\altaffilmark{3,4}}
\author{David J. Schlegel\altaffilmark{2}}

\altaffiltext{1}{Berkeley Center for Cosmological Physics, Berkeley, CA 94720, 
USA}
\altaffiltext{2}{Lawrence Berkeley National Laboratory, Berkeley, CA, 94720, 
USA}
\altaffiltext{3}{Department of Astronomy \& Astrophysics and Dunlap Institute, 
University of Toronto, Toronto, ON M5S 3H4, Canada}
\altaffiltext{4}{Department of Physics \& Astronomy, University of Waterloo,
200 University Avenue West, Waterloo, ON, N2L 3G1, Canada}

\begin{abstract} 
The Near Earth Object Wide-field Infrared Survey Explorer (NEOWISE)
Reactivation mission released data from its first full year of observations
in 2015. This data set includes $\sim$2.5 million exposures in each of W1 and
W2, effectively doubling the amount of WISE imaging available at 3.4$\mu$m and 
4.6$\mu$m relative to the AllWISE release. We have created the first ever 
full-sky set of coadds combining all publicly available W1 and W2 exposures 
from both the AllWISE and NEOWISE-Reactivation (NEOWISER) mission phases. We 
employ an adaptation of the unWISE image coaddition framework \citep{lang14}, 
which preserves the native WISE angular resolution and is optimized for forced 
photometry. By incorporating two additional scans of the entire sky, we not 
only improve the W1/W2 depths, but also largely eliminate time-dependent 
artifacts such as off-axis scattered moonlight. We anticipate that our new 
coadds will have a broad range of applications, including target selection for 
upcoming spectroscopic cosmology surveys, identification of distant/massive 
galaxy clusters, and discovery of high-redshift quasars. In particular, our 
full-depth AllWISE+NEOWISER coadds will be an important input for the Dark 
Energy Spectroscopic Instrument (DESI) selection of luminous red galaxy and 
quasar targets. Our full-depth W1/W2 coadds are already in use within the 
DECam Legacy Survey (DECaLS) and Mayall z-band Legacy Survey (MzLS) reduction 
pipelines. Much more work still remains in order to fully leverage NEOWISER 
imaging for astrophysical applications beyond the solar system.
\end{abstract}  
 

\section{Introduction}

The Wide-field Infrared Survey Explorer \cite[WISE; ][]{wright10} has performed
a full-sky imaging survey in four broad mid-infrared bandpasses centered at 
3.4, 4.6, 12 and 22 microns, labeled W1-W4 from blue to red. WISE has 
dramatically enhanced our knowledge of the mid-infrared sky, and publicly 
released numerous catalog and imaging data products of high value to the 
astronomical community.

WISE launched in December 2009, and undertook a seven 
month, full-sky survey in all of W1-W4 from February 2010 through August
2010. In September 2010, the solid hydrogen cryogen used to cool the W3 and W4 
instrumentation was depleted, significantly reducing the quality of W3 
imaging and rendering W4 unusable. Nevertheless, WISE continued surveying 
the sky from September 2010 to February 2011 in W1 and W2, including 
a portion of the mission referred to as NEOWISE \citep{neowise}. In February 
2011 WISE was placed in hibernation. In October 2013, WISE was reactivated, and
recommenced surveying the sky in W1 and W2. This W1/W2 survey is referred to as 
NEOWISE-Reactivation (NEOWISER) and is expected to continue until 2017. The
first-year NEOWISER data products, including all single-exposure images,
were publicly released in March 2015. Importantly, the NEOWISER images are of 
very nearly the same high quality as those of the pre-hibernation WISE mission
\citep{neowiser}.

Several data products consisting of full-sky, stacked WISE imaging are 
currently available for the first 13 months of data. The WISE team has created 
a set of ``Atlas'' coadds smoothed by the point spread function (PSF) using the
first 7 months of data (the All-Sky release), and the first 13 months of data 
(the AllWISE release). In an independent processing effort, \cite{lang14} has 
produced custom ``unWISE'' stacks analogous to the AllWISE Atlas images, but at
the full spatial resolution of the instrument. These unWISE stacks are 
optimized for forced photometry, and have proven to be an important input for 
eBOSS target selection \citep{lang14b, eboss_qso, eboss_lrg}.

However, until now, no full-sky set of W1/W2 coadds combining all pre and post 
reactivation exposures has existed. The primary motivation for such a
data product is the enhanced depth achieved relative to AllWISE-only coadds. 
Among other benefits, this added depth will improve the utility of WISE for 
selecting higher-redshift spectroscopic targets, in particular for the upcoming
Dark Energy Spectroscopic Instrument \citep[DESI,][]{desi}. Furthermore, 
folding in two additional scans of WISE data at each sky location allows
time-dependent artifacts to be nulled, largely eliminating spatial 
nonuniformities in image quality and derivative catalogs.

% add sentence to above paragraph about Faherty arxiv survey saying
% WISE+NEOWISER coadds are highly valuable ?

Here we present a new set of full-sky coadds generated by
combining all publicly available W1/W2 exposures from the AllWISE and NEOWISER
programs, using an adaptation of the \cite{lang14} unWISE methodology. These 
`full-depth' coadds are publicly available online.\footnote{http://unwise.me}

In $\S$\ref{sec:data} we briefly describe the W1/W2 single-exposure data set 
from which our coadds are constructed. In $\S$\ref{sec:coadd}, we review the 
important aspects of the unWISE coaddition framework we employ and list the 
processing features which are newly introduced in this work. In 
$\S$\ref{sec:calib} we describe an empirical photometric calibration we
derived in order to combine pre and post reactivation WISE images. In 
$\S$\ref{sec:moon} we describe our rejection of time-dependent artifacts, 
particularly scattered moonlight. In $\S$\ref{sec:recover} we describe
our procedure for recovering Moon-contaminated exposures. In 
$\S$\ref{sec:results} we highlight some important aspects of the full-sky set 
of full-depth coadds generated by our processing. In $\S$\ref{sec:depth} we 
present a catalog-level validation of the improvements in WISE depth
that result from doubling the amount of W1/W2 imaging. We conclude in 
$\S$\ref{sec:future} with a brief discussion of the work which still remains to
be done with existing and future NEOWISER imaging.

\section{Data}
\label{sec:data}

The WISE single-exposure ``L1b'' images represent
the input data for our W1/W2 coadds. Specifically, for each L1b frameset, we 
make use of the per-band \verb|-int-|, \verb|-msk-| and \verb|-unc-| images, 
which respectively give the measured sky intensity and associated per-pixel 
bitmask values and uncertainty estimates. We have obtained a local copy of 
these files for every publicly available frameset, including those 
from the AllWISE release ($\sim$2.8M framesets) and first-year NEOWISER release
($\sim$2.5M framesets). In all we have analyzed $\sim$5.3M framesets in each of
W1 and W2, corresponding to a total of $\sim$32M L1b image files, $\sim$71TB of 
input image data and $\sim$33$\times$10$^{12}$ pixels.

In addition to L1b images, we also make use of several catalog-level WISE
data products. During the photometric calibration described in 
$\S$\ref{sec:calib}, we select sources to photometer based on the AllWISE 
Source Catalog \citep{cutri13}. Also, to flag and reject bright solar system 
planets ($\S$\ref{sec:moon}), we employ the WISE Known Solar System Object
Possible Association List for each mission phase 
\citep{cutri12, cutri13,cutri15}.

\section{Image Coaddition Methodology}
\label{sec:coadd}

To stack the W1/W2 single exposures, we make use of the \cite{lang14}
unWISE coaddition framework, and perform our image processing with an 
adaptation of the codebase from that work. We briefly mention
a few of the salient aspects of the unWISE coaddition methodology here; for
a full discussion see \cite{lang14}.

Like the official WISE Atlas coadds, unWISE processing divides the sky
into a set of 18,240 1.56$^{\circ}$$\times$1.56$^{\circ}$ tiles
arranged along iso-declination rings. Whereas the Atlas images
are smoothed by the WISE PSF, the unWISE code uses Lanczos interpolation to 
preserve the native WISE angular resolution during coaddition, creating stacked
outputs which are 2048 pixels on a side, with 2.75$''$ pixels.

During the course of this work, various modifications have been made to the
 unWISE codebase and methodology. Here we highlight the important 
updates/changes:

\begin{itemize}
\item We include all publicly available NEOWISER W1 and W2 exposures, 
approximately doubling the number of input L1b frames relative to the 
\cite{lang14} processing.
\item We adopt custom zero points based on repeat photometry at the ecliptic 
poles. In contrast, the \cite{lang14} unWISE processing adopted zero points 
extracted from the L1b header metadata.
\item We explicitly reject exposures contaminated by the Moon and/or solar 
system planets. No such rejection was included in the \cite{lang14} unWISE 
W1/W2 coadds, although these and other artifacts were addressed to some extent 
via general-purpose outlier rejection.
\item In this work we attempt to recover Moon-affected frames by applying 
polynomial background level corrections to the contaminated exposures.
\end{itemize}

The outputs generated in this work follow the same data model as those of 
\cite{lang14}. For each tile, a stacked intensity image is created, as well as
auxiliary maps of useful quantities such as the per-pixel inverse variance and
integer coverage. Like those of \cite{lang14}, our coadds have units of Vega 
nanomaggies.

\section{Custom Photometric Calibration}
\label{sec:calib}

In order to combine frames across multiple mission phases, it is necessary
to place all exposures on a common photometric calibration so that
the multiplicative scalings of all input images are consistent. Each L1b
image includes a \verb|MAGZP| header keyword which gives the nominal
Vega zero point of that exposure. These zero points are essentially
predictions of system throughput based on predictors such as beam splitter
assembly temperature, and have been found to differ by up to several percent 
relative to zero point variations measured empirically with single-exposure
photometry of calibrator sources \citep[e.g.][$\S$V.3.a.iii.1]{cutri13}.

We therefore sought to derive an empirical relative photometric calibration
across all mission phases accurate at the sub-percent level. To do so, we 
analyzed repeat measurements of compact sources near the 
ecliptic poles, where WISE has gathered data every $\sim$$95$ minutes 
throughout the entire mission. Specifically, our sample consists of moderately 
bright, unsaturated compact sources with $|\beta| > 85^{\circ}$, avoiding a 
wedge defined by $-90^{\circ}$$<$$\lambda$$<25^{\circ}$ near the south ecliptic
pole to exclude the LMC. The positions and average magnitudes of our 
moderately bright source sample were drawn from the AllWISE Source Catalog, 
selecting W1 sources with 10.6 $<$ \verb|w1mpro| $<$ 13.1 and W2 sources
with 9.2 $<$ \verb|w2mpro| $<$ 11.7, always requiring \verb|w?cc_map|=0 in the 
band of interest.

These spatial, magnitude and flag cuts yield samples of $\sim$109,000 
($\sim$27,000) unique calibrator sources in W1 (W2). We perform aperture 
photometry using \verb|djs_phot| with a 27.5$''$ radius for each calibrator 
source in every L1b exposure in which it appears sufficiently far from the 
image boundary. This results in a catalog of $\sim$45M ($\sim$15M) W1 (W2) 
single-epoch aperture fluxes (in units of DN), which will form the basis for 
our derived time-dependent zero points. The typical calibrator source 
contributes $\gtrsim$400 epochs of photometry.

% consider dropping last sentence ?

\begin{figure*} [ht]
 \begin{center}
  \epsfig{file=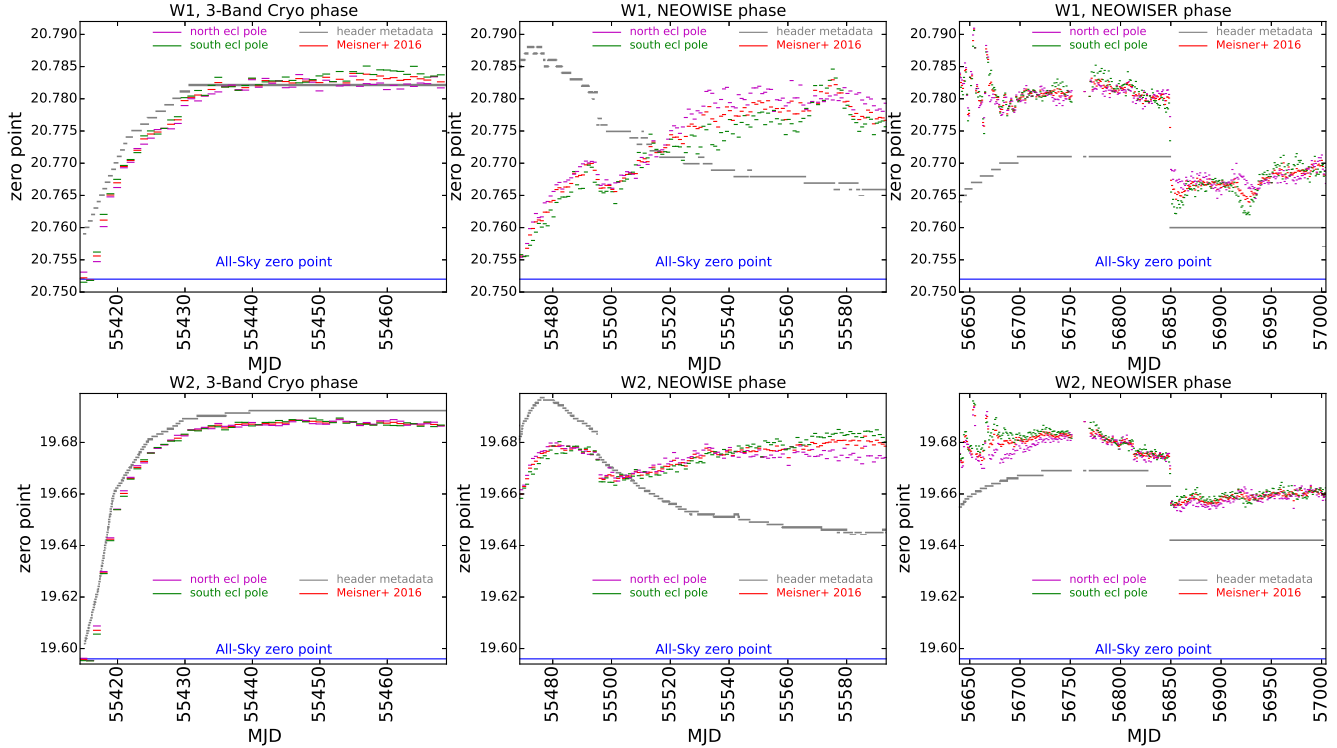, width=7.0in}
  \caption{\label{fig:zp} Summary of our custom photometric calibration. In 
           each panel, the red dashes give the per-day
           photometric zero points we derived based on repeat measurements
           of stars near the ecliptic poles, as described in 
           $\S$\ref{sec:calib}. The gray lines show the
           \texttt{MAGZP} zero point values provided by the L1b header 
           metadata. The magenta (green) dashes
           represent the zero point time trends which would have resulted had
           we restricted our analysis to aperture photometry of only the north 
           (south) ecliptic pole.}
 \end{center}
\end{figure*}

We desired a photometic calibration with time resolution of one day. To
achieve this, we grouped our single-exposure aperture photometry measurements 
in two ways. First, for each 
unique source, we lumped all of its All-Sky phase measurements together to 
obtain its median flux (DN) in our aperture photometry system during this 
phase. This is justified because the All-Sky release photometric zero points 
are known to be remarkably stable \citep{jarrett11}. Indeed, using our ecliptic
pole single-exposure photometry database, we were able to confirm that the 
All-Sky phase zero point was stable at the $\le$2 mmag level in both bands. 
Next, for every aperture flux after the All-Sky phase, we calculated a 
multiplicative enhancement factor
implied by the ratio of that measurement to the appropriate source's
median All-Sky phase flux. We then grouped these 
flux enhancement factors into one-day bins, and quote the median per 
bin as the change in multiplicative
image scaling relative to the All-Sky zero point. For the  
All-Sky phase zero points, we adopted the \verb|MAGZP| values of 20.752 in W1 
and 19.596 in W2.

Figure \ref{fig:zp} shows our derived zero points for each WISE mission phase 
as compared to the \verb|MAGZP| values obtained from the L1b headers.  
In general, our per-day zero points agree reasonably
well with the \verb|MAGZP| values, although there are often 
percent-level differences, and at times disagreement at the several percent
level. In some cases the empirically measured time trends within a particular 
mission phase show qualitative disagreement with the header metadata (e.g. both
the W1 and W2 zero points during the NEOWISE mission phase). When scaling each
L1b image according to its zero point during coaddition, we employ an 
interpolation scheme meant to avoid directly using our somewhat noisy per-day 
measurements. Specifically, we create a smooth approximate representation of 
the measured per-day zero point time-series, based on a series of polynomials 
and error functions, tapering between segments such that the resulting curve is
smooth.

\section{Removing Time-dependent Artifacts}
\label{sec:moon}

The WISE scan strategy is such that a typical sky location will be 
observed at approximately six month intervals, with each six-monthly
``visit'' yielding a series of $\sim$12 exposures over a $\sim$1
day time span. Within the AllWISE release, most of the sky contains just two 
visits
of W1/W2 imaging. Incorporating exposures from both the AllWISE and
NEOWISER phases effectively doubles this value to four visits everywhere
on the sky. If we coadded the AllWISE+NEOWISER data naively, without
concern for time dependent artifacts, we would risk corrupting regions of the 
sky that were pristine during the AllWISE phase. Instead, we have found that by 
leveraging the added redundancy of extra NEOWISER visits while carefully 
addressing time-dependent artifacts, we can create full-depth coadds which are 
nearly artifact-free over the entire sky.

The dominant time-dependent artifact in W1/W2 images is off-axis
scattered light from the Moon, which can significantly contaminate
images at angular separations of up to many tens of degrees. This
contamination manifests itself in L1b exposures as a strongly spatially 
variable background level, which in certain images can be smooth, but in
others can show a very complex morphology.

\begin{figure*} [ht]
 \begin{center}
  \epsfig{file=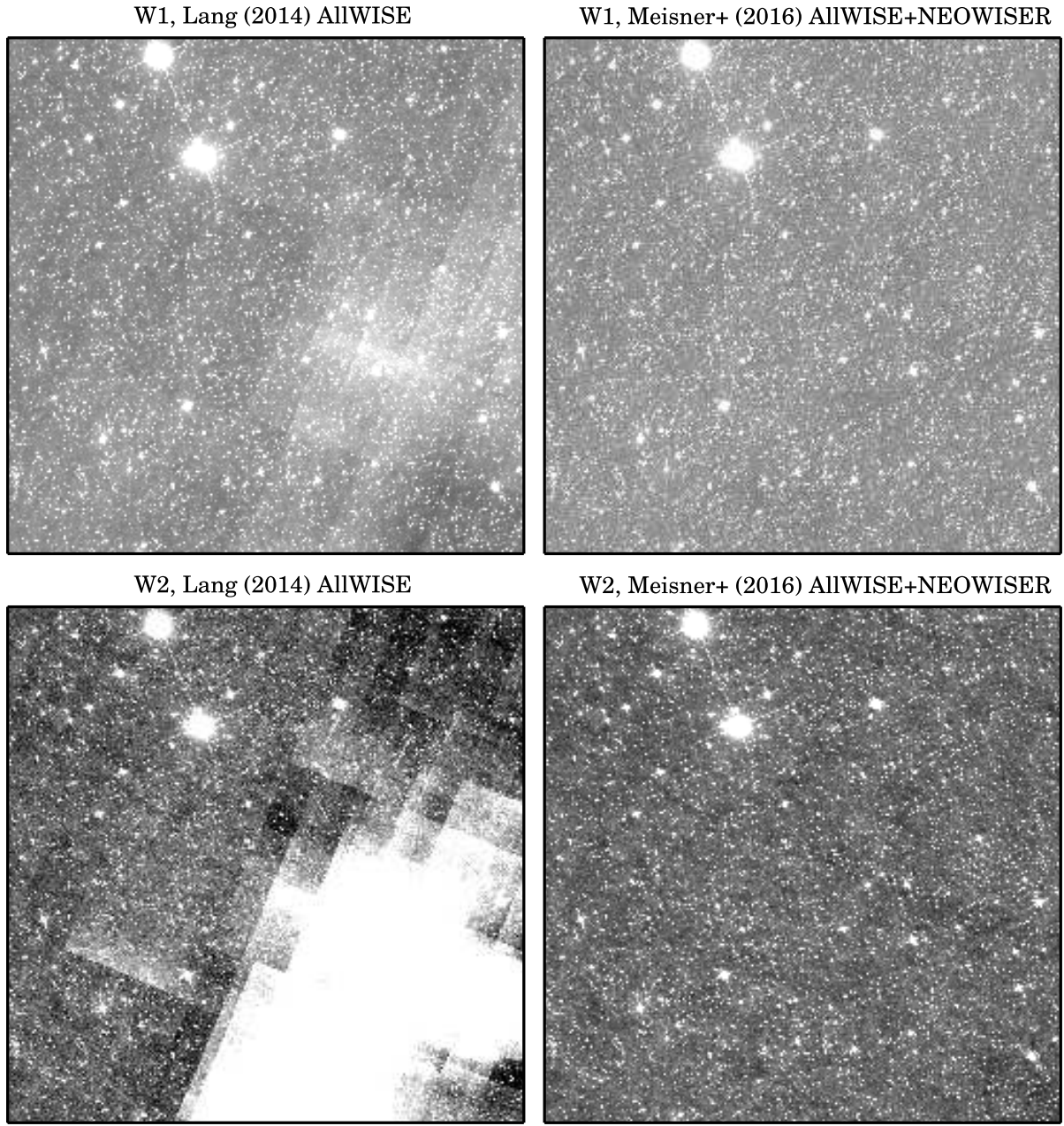, width=6.5in}
  \caption{\label{fig:moon_3433p000} Tile 3433p000, a 
           1.56$^{\circ}$$\times$1.56$^{\circ}$ region of sky at low
           ecliptic latitude which experienced strong Moon contamination 
           during the AllWISE mission phase. Rejecting Moon-contaminated
           frames as described in $\S$\ref{sec:moon} while including
           two extra visits worth of NEOWISER data has dramatically
           improved the background level uniformity relative to the
           \cite{lang14} version of this tile. For this tile, we were
           able to recover 58\% (18\%) of Moon-contaminated frames in W1 (W2).}
 \end{center}
\end{figure*}

In the \cite{lang14} unWISE processing, no steps were taken specifically
to mitigate scattered moonlight in W1 and W2. However, that analysis did 
address Moon contamination in W3 and W4. \cite{lang14}
inspected all W3/W4 exposures flagged with the \verb|MOON_MASKED| bit, and 
discarded those frames with abnormally large pixel value standard deviations,
indicative of a strongly varying background level (see \citealt{lang14} $\S$2
for full details). In the present work we have applied this same
Moon rejection criterion to W1 and W2 frames. Because of the added redundancy
of two extra NEOWISER scans, we are thus able to reject many W1/W2
frames which are contaminated by moonlight, while still retaining 
sufficient artifact-free coverage everywhere on the sky to avoid leaving
any holes in the stacks.

Figure \ref{fig:moon_3433p000} shows the dramatic improvement achieved toward 
maintaining a consistent coadd background level by virtue of folding in 
NEOWISER data and applying the frame-level Moon rejection cut, for a tile with 
severe Moon contamination during the AllWISE phase.

A second, less common type of time-dependent artifact results when
bright solar system planets (Mars, Jupiter and Saturn) pass through
the WISE field of view. These planet sightings are prominent in the 
\cite{lang14} unWISE stacks, as no steps were taken to address such
occurrences. In constructing our new full-depth coadds, we have used the Known 
Solar System Possible Association List to identify all exposures in which
Mars, Jupiter or Saturn fall within the WISE field of view. We discard
such frames completely during coaddition and make no attempt to recover them.

Bright planets are also accompanied by scattered light halos a few degrees in 
size. Therefore, we additionally use ephemerides to identify all frames within 
2.5$^{\circ}$ of these planets. During coaddition,
such frames are initially ignored. However, we later attempt to recover such
frames according to the procedure described below in $\S$\ref{sec:recover}.

\section{Recovering Contaminated Frames}
\label{sec:recover}

% quadrant warping example figure

\begin{figure*}[ht]
 \begin{center}
  \epsfig{file=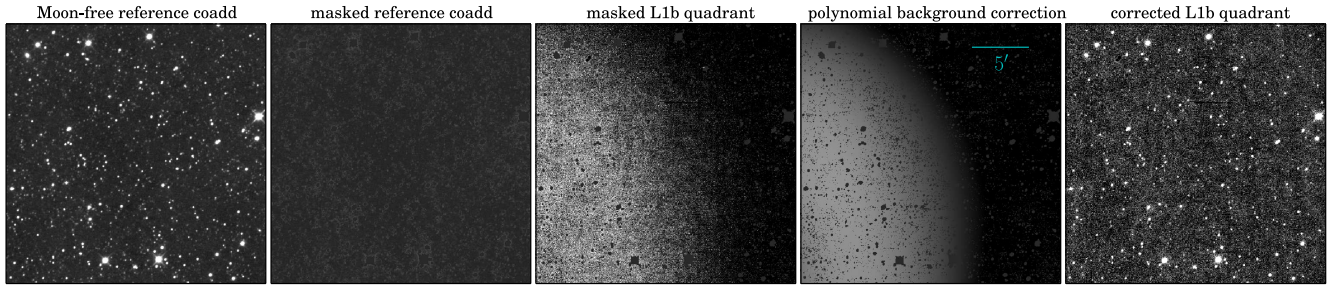, width=7.0in}
  \caption{\label{fig:warp_example} Illustration of the procedure by which
           we recover Moon-contaminated exposures, as described in 
           $\S$\ref{sec:recover}. Shown here is quadrant 2
           of W1 exposure 05245b140. This quadrant was successfully recovered.
           The polynomial background correction is subtracted
           from every pixel in the L1b quadrant, but is shown masked
           for the sake of comparison to the masked L1b quadrant.}
% mention which tile used as reference?
 \end{center}
\end{figure*}

Although we are able to dramatically reduce the impact of Moon contamination
on our coadds with the exposure rejection procedure of $\S$\ref{sec:moon},
we would ideally like to recover as much Moon-contaminated data as possible,
rather than simply discard it all outright. To that end, we have added an 
after-burner step to our coaddition procedure, during which we attempt to 
salvage frames that were flagged with \verb|MOON_MASKED| and displayed 
abnormally large pixel value standard deviations. The procedure we employ is a 
variant of that described in $\S6.4.1$ of \cite{meisner14}.

The first two rounds of unWISE coaddition still proceed exactly as 
described in \cite{lang14}. These steps yield a Moon-free stack which we 
subsequenty use as a reference image to compare against, and derive low-order 
corrections for, each Moon-contaminated frame.

For each frame initially rejected on the basis of Moon contamination, we 
first resample the exposure onto the coadd astrometry. We then divide
the exposure into quadrants which we analyze separately. For each quadrant,
we will attempt to model the Moon contamination with a polynomial 
offset as a function of L1b $x$, $y$ pixel coordinates. We begin by masking out 
the brightest and faintest 5\% of pixels in the reference coadd, since pixels
with bright compact sources will not be very informative for background level
modeling. We then fit the difference between the masked L1b quadrant and masked
reference coadd with a fourth order polynomial in L1b $x$, $y$ coordinates. We 
evaluate the chi-squared of this model, using the reference coadd's per-pixel
standard deviation values to construct per-pixel uncertainty estimates.

For each quadrant, we deem the polynomial correction to be a satisfactory 
description of the scattered moonlight if the mean per-pixel chi-squared is 
less than 2.5. In that case, we then subtract the polynomial correction from 
the quadrant, and consider the quadrant ``recovered''. Quadrants with poor 
chi-squared are discarded and remain excluded from the coadd. Once a list of 
all recovered quadrants has been assembled, these are accumulated into the 
existing reference coadd to produce a final set of outputs for the tile under 
consideration.

Figure \ref{fig:warp_example} provides an illustration of our polynomial 
background modeling procedure applied to a single L1b quadrant. We were able to 
recover 54\% of Moon-contaminated data in W1 and 33\% in W2. We also apply this
polynomial correction procedure to frames that we flagged as potentially 
affected by scattered light halos from bright solar system planets. We
categorically exclude \verb|qual_frame|=0 exposures, making no attempt to 
recover such frames.

\section{Overview of Results}
\label{sec:results}

\begin{figure*} [ht]
 \begin{center}
  \epsfig{file=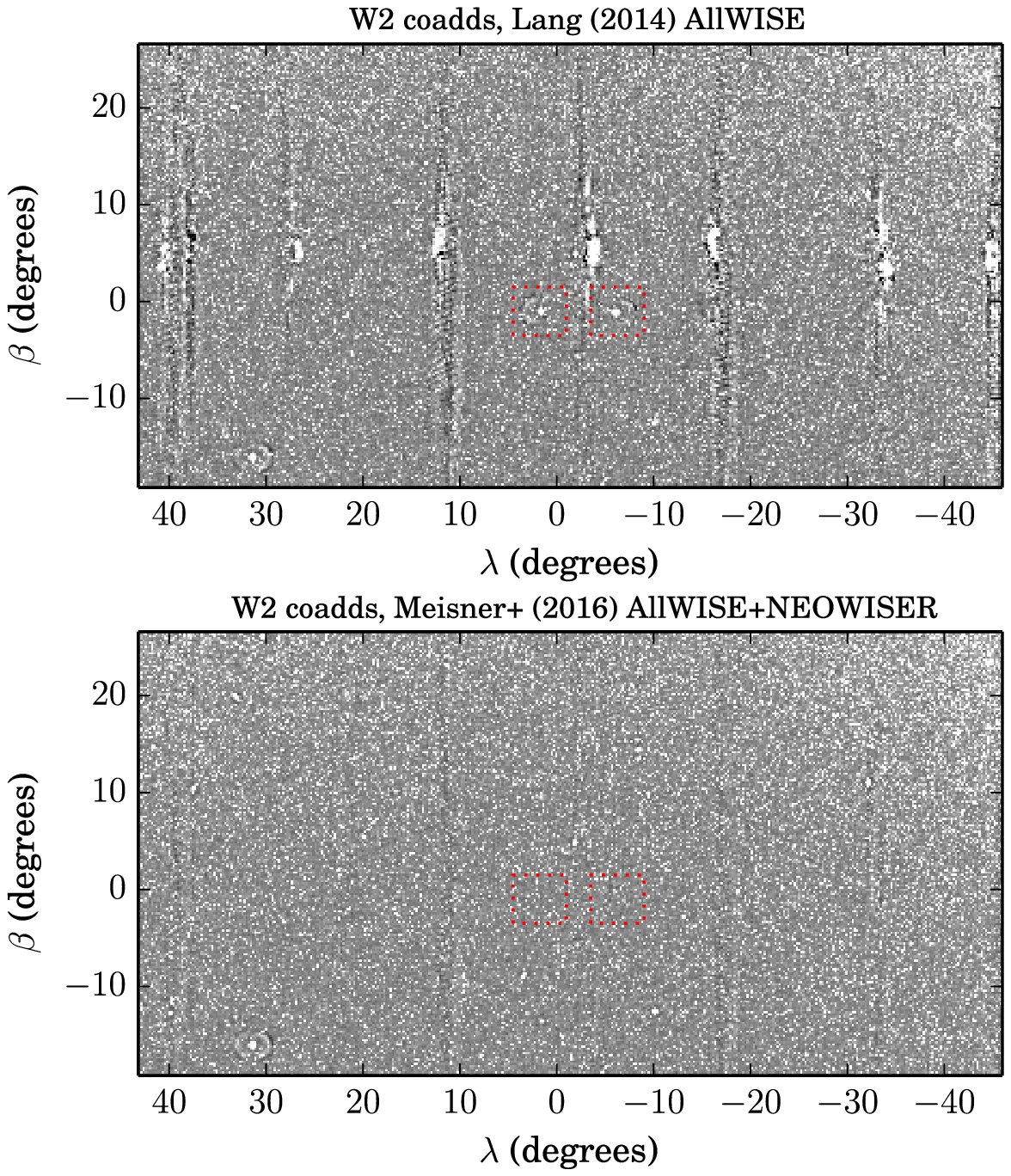, width=7.0in}
  \caption{\label{fig:w2_large} A large portion of the south Galactic
           cap near the ecliptic plane is shown in W2. Top: \cite{lang14}
           unWISE coadds based on the AllWISE release imaging and without
           rejection of Moon-contaminated frames. Bottom: same region of
           sky in our new AllWISE+NEOWISER stacks, with double the
           redundancy in sky coverage and rejection/recovery of 
           Moon-contaminated frames. It is clear that the scattered
           moonlight, which appears as a series of vertical streaks, has been 
           largely removed in W2, although some traces still remain. The
           two dotted red boxes show locations where Jupiter passed through
           the WISE field of view. The imprints of such planet sightings have 
           now been removed.}
 \end{center}
\end{figure*}

\begin{figure*} [ht]
 \begin{center}
  \epsfig{file=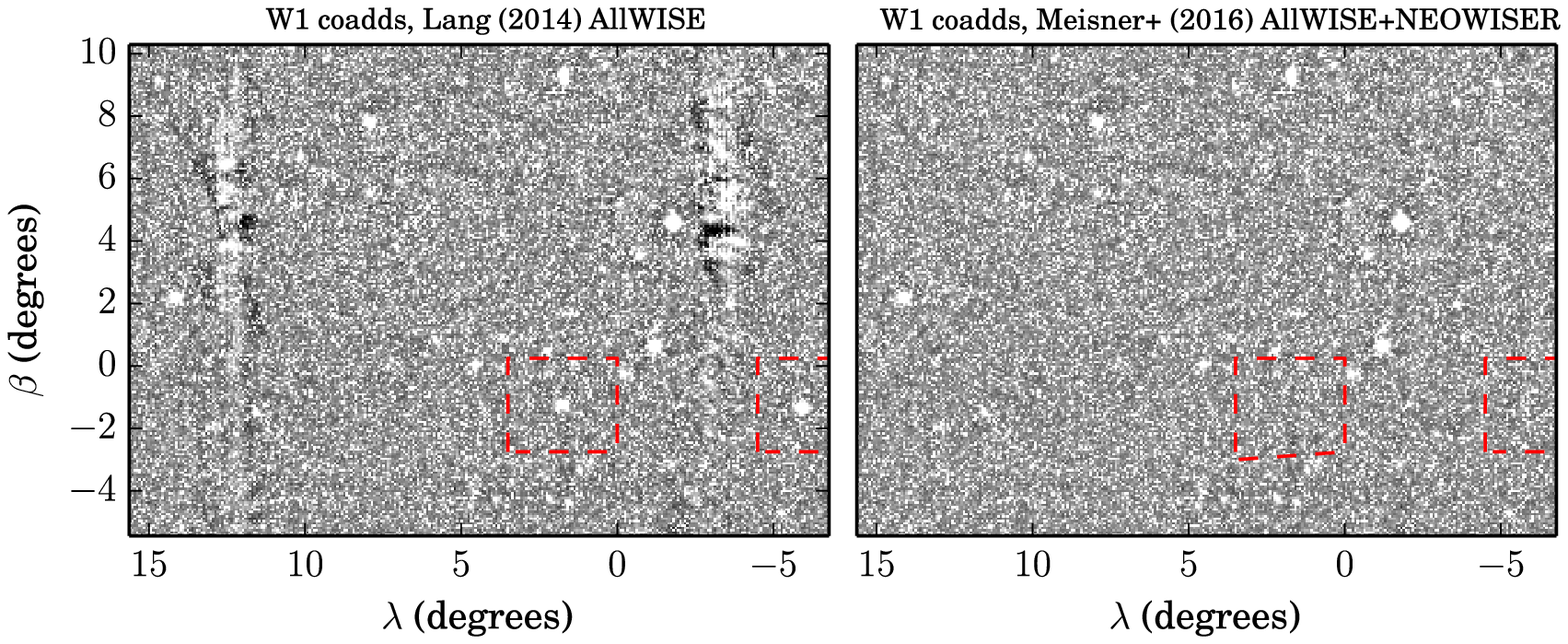, width=7.0in}
  \caption{\label{fig:w1_large} A subsection of the region displayed in Figure
           \ref{fig:w2_large}, now shown in W1. Left: \cite{lang14}
           unWISE coadds based on the AllWISE release imaging and without
           rejection of Moon-contaminated frames. Right: same region of
           sky in our new AllWISE+NEOWISER stacks, with double the
           redundancy in sky coverage and rejection/recovery of 
           Moon-contaminated frames. Scattered moonlight (vertical streaks
           in left panel) has been completely eliminated in W1. The
           two dashed red boxes show locations where Jupiter passed through
           the WISE field of view. The imprints of such planet sightings have 
           now been removed}
 \end{center}
\end{figure*}

Figures \ref{fig:w2_large} and \ref{fig:w1_large} show large-scale renderings 
of our full-depth coadds over portions of south Galactic cap. It is apparent 
that the Moon contamination has been completely eliminated in
W1 and dramatically reduced in W2. It is possible that simply folding in 
additional NEOWISER W2 frames from forthcoming data releases will diminish the 
remaining Moon imprint to such an extent that no additional processing 
modifications will be required to address this issue.

\begin{figure*}[ht]
 \begin{center}
  \epsfig{file=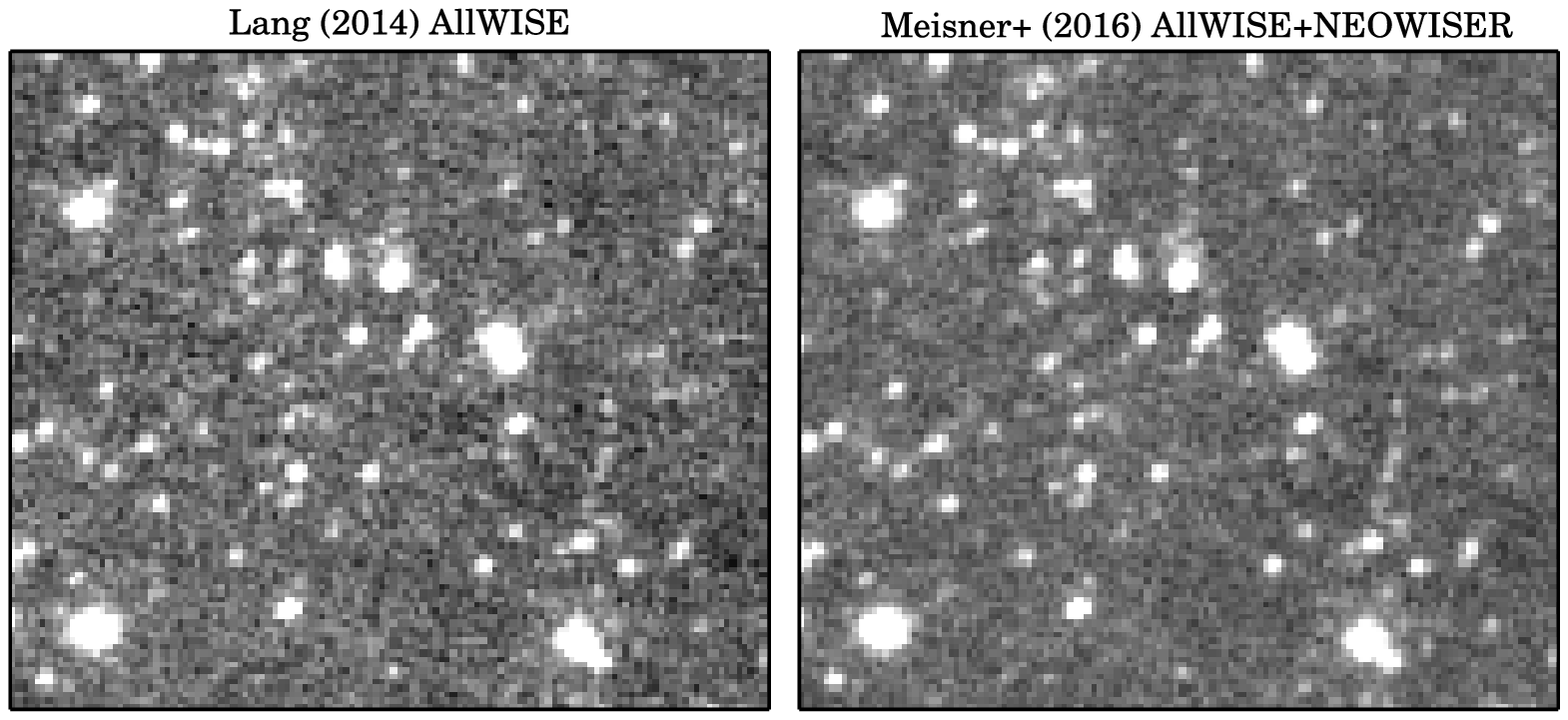, width=6.0in}
  \caption{\label{fig:image_noise} Example of decreased image noise
           in W1 for a $\sim$6$'$ cutout extracted from tile 3353m107, centered
           at ($\alpha$, $\delta$) = (335.4585$^{\circ}$, $-$10.5801$^{\circ}$). 
           This is a representative location at low ecliptic latitude and high 
           Galactic latitude,  with $\sim$2$\times$ as much coverage in our 
           AllWISE+NEOWISER coadd relative to that of \cite{lang14}. Both
           images are rendered with identical grayscale stretches.}
 \end{center}
\end{figure*}

Figure \ref{fig:image_noise} shows zoom-ins of a low ecliptic latitude field,
illustrating visually the reduction in statistical noise which has been 
achieved by doubling the number of input exposures.

% figure showing close-up at low ecl latitude with reduced noise

Given that we imposed various frame-level cuts to eliminate time-dependent 
artifacts, it is reasonable to ask whether we have created any zero-coverage 
holes as a result. We have checked all 18,240 of our integer coverage maps, and
find that every pixel has $\ge$18 epochs in W1 and $\ge$15 epochs in W2.

\section{Catalog-level Validation}
\label{sec:depth}

It is important to quantify the effect of increased coverage on
the W1/W2 depths achieved by catalogs based on our AllWISE+NEOWISER coadds 
relative to those based on the AllWISE-only coadds of \cite{lang14}. A
complete characterization should systematically explore a range of 
ecliptic and Galactic latitudes, to consider the full spectrum of interplay
between decreased statistical noise and confusion. For isolated sources in the 
sky noise limit, we would expect a doubling of the W1/W2 imaging data
to increase the depth in each band by 0.38 mag.

The only existing catalogs based on our AllWISE+NEOWISER coadds
are those from DR2 of the DECam Legacy Survey\footnote{http://legacysurvey.org} 
\citep[DECaLS,][]{decals}, which include W1-W4 forced photometry for all
optically detected sources. On the other hand, optical sources in DECaLS DR1
are accompanied by forced photometry of the \cite{lang14} unWISE coadds. 
Therefore, comparing the uncertainties on W1/W2 fluxes in DR2 versus DR1 can 
allow us to obtain estimates of the forced-photometry depth increases for 
regions of low ecliptic and high Galactic latitude.

To make such a comparison, for each of W1 and W2, we select a sample of 
isolated DECaLS DR2 sources with \verb|type|=`PSF', \verb|wise_fracflux|$<$0.1,
more than one DECam observation, and WISE S/N within the range 10$\pm$1. We
then obtain a DR1-DR2 comparison sample by cross-matching these DR2 objects
with DR1 \verb|type|=`PSF' sources, using a 1$''$ matching radius. The resulting
comparison samples have $\sim$415,000 objects. For such
sources, we find median reductions in forced photometry flux uncertainties
of 1.38$\times$ (1.30$\times$) in W1 (W2). The corresponding median increase in
coverage is 1.9$\times$ in both bands, leading us to expect a factor of 
1.38$\times$ decrease in flux uncertainties based purely on reduced statistical 
noise.

These values are not intended to represent the enhanced
depths that would result from comparing WISE-only catalogs which begin with
a source detection step on the WISE coadds. Since the DR2 forced photometry 
does not attempt to account for faint, infrared-only sources which are
newly revealed in the AllWISE+NEOWISER stacks, the photometric uncertainty
reduction factors we have derived serve mostly as quantitative confirmation 
of the decreased statistical noise shown in Figure \ref{fig:image_noise}.

Much more work is needed to characterize the increased sensitivity
of WISE to faint infrared-only sources due to the inclusion of NEOWISER
imaging, as this will inform forecasts about the final 3.4$\mu$m and 
4.6$\mu$m depths expected upon WISE's permanent retirement.

\section{Conclusion \& Future Work}
\label{sec:future}

We have created a full-sky set of W1/W2 coadds which combine all publicly
available exposures from both the AllWISE and NEOWISER releases. Doubling the 
amount of WISE imaging relative to the AllWISE release has resulted in improved
W1/W2 depths and allowed for the elimination of nearly all time-dependent 
artifacts. Our new AllWISE+NEOWISER W1/W2 coadds are publicly available
via http://unwise.me.

% could list future improvement's i'd like to implement for NEO(2+) coadds
% for instance masking of bright star ghosts and correction
% of bright star scattered light, possibly creation of a bitmask output
% like that of meisner/finkbeiner dust coadds
% also min/max removal step as added protection against outliers

Although the present analysis constituted a significant data processing 
endeavor, it represents only a small fraction of the work that must
be done to maximize science return from the NEOWISER imaging data set.
Creation of time-resolved coadds spanning the $\sim$5 year AllWISE-NEOWISER
time baseline would enable a wealth of important time-domain projects,
ranging from brown dwarf searches to infrared quasar variability studies. As
additional years of NEOWISER data become publicly available, it will be 
necessary to continue updating the full-depth AllWISE+NEOWISER coadds to 
maximize the achieved W1/W2 depths. Finally, the creation of WISE-only (as 
opposed to forced photometry) catalogs based on deep coadds which incorporate
NEOWISER images will be needed to enable a variety of exciting discoveries.

\acknowledgments{
We thank Roc Cutri for his guidance in making use of the WISE/NEOWISE data 
products. We thank Stephen Bailey for downloading the L1b images to NERSC. We 
thank Doug Finkbeiner for providing access to the big-memory computers which
were used to run coadds near the ecliptic poles. We thank Peter Nugent for 
data management and HPC assistance. We thank Ben Weaver, Debbie Bard and Rollin
Thomas for additional HPC advice. We thank many members of the DECaLS/MzLS 
data team for valuable feedback, especially Arjun Dey, Doug Finkbeiner, John 
Moustakas, Bob Blum and Peter Eisenhardt.

This research is supported by the Director, Office of Science, Office of High 
Energy Physics of the U.S. Department of Energy under Contract No. 
DE–AC02–05CH1123, and by the National Energy Research Scientific Computing 
Center, a DOE Office of Science User Facility under the same contract; 
additional support for DESI is provided by the U.S. National Science
Foundation, Division of Astronomical Sciences under Contract No. AST-0950945 to
the National Optical Astronomy Observatory; the Science and Technologies 
Facilities Council of the United Kingdom; the Gordon and Betty Moore 
Foundation; the Heising-Simons Foundation; and by the DESI Member Institutions.
The Dunlap Institute is funded through an endowment established by the David 
Dunlap family and the University of Toronto.

This research made use of the NASA Astrophysics Data System (ADS) and the IDL 
Astronomy User's Library at Goddard.\footnote{Available at 
\texttt{http://idlastro.gsfc.nasa.gov}}

This research makes use of data products from the Wide-field Infrared
Survey Explorer, which is a joint project of the University of California, Los
Angeles, and the Jet Propulsion Laboratory/California Institute of Technology,
funded by the National Aeronautics and Space Administration. This research
also makes use of data products from NEOWISE, which is a project of the Jet
Propulsion Laboratory/California Institute of Technology, funded by the
Planetary Science Division of the National Aeronautics and Space
Administration. }

\bibliographystyle{apj}
\bibliography{fulldepth_neo1.bib}

\begin{thebibliography}{14}
\expandafter\ifx\csname natexlab\endcsname\relax\def\natexlab#1{#1}\fi

\bibitem[{{Cutri} {et~al.}(2015){Cutri}, {Mainzer}, {Conrow}, {Masci}, {Bauer},
  {Dailey}, {Kirkpatrick}, {Fajardo-Acosta}, {Gelino}, {Grillmair}, {Wheelock},
  {Yan}, {Harbut}, {Beck}, {Wittman}, {Wright}, {Masiero}, {Grav}, {Sonnett},
  {Nugent}, {Kramer}, {Stevenson}, {Eisenhardt}, {Fabinsky}, {Tholen}, {Papin},
  {Fowler}, \& {McCallon}}]{cutri15}
{Cutri}, R.~M., {Mainzer}, A., {Conrow}, T.,  {et~al.} 2015, {Explanatory
  Supplement to the NEOWISE Data Release Products}, Tech. rep.

\bibitem[{{Cutri} {et~al.}(2012){Cutri}, {Wright}, {Conrow}, {Bauer},
  {Benford}, {Brandenburg}, {Dailey}, {Eisenhardt}, {Evans}, {Fajardo-Acosta},
  {Fowler}, {Gelino}, {Grillmair}, {Harbut}, {Hoffman}, {Jarrett},
  {Kirkpatrick}, {Leisawitz}, {Liu}, {Mainzer}, {Marsh}, {Masci}, {McCallon},
  {Padgett}, {Ressler}, {Royer}, {Skrutskie}, {Stanford}, {Wyatt}, {Tholen},
  {Tsai}, {Wachter}, {Wheelock}, {Yan}, {Alles}, {Beck}, {Grav}, {Masiero},
  {McCollum}, {McGehee}, {Papin}, \& {Wittman}}]{cutri12}
{Cutri}, R.~M., {Wright}, E.~L., {Conrow}, T.,  {et~al.} 2012, {Explanatory
  Supplement to the WISE All-Sky Data Release Products}, Tech. rep.

\bibitem[{{Cutri} {et~al.}(2013){Cutri}, {Wright}, {Conrow}, {Fowler},
  {Eisenhardt}, {Grillmair}, {Kirkpatrick}, {Masci}, {McCallon}, {Wheelock},
  {Fajardo-Acosta}, {Yan}, {Benford}, {Harbut}, {Jarrett}, {Lake}, {Leisawitz},
  {Ressler}, {Stanford}, {Tsai}, {Liu}, {Helou}, {Mainzer}, {Gettings},
  {Gonzalez}, {Hoffman}, {Marsh}, {Padgett}, {Skrutskie}, {Beck}, {Papin}, \&
  {Wittman}}]{cutri13}
---. 2013, {Explanatory Supplement to the AllWISE Data Release Products}, Tech.
  rep.

\bibitem[{{Jarrett} {et~al.}(2011){Jarrett}, {Cohen}, {Masci}, {Wright},
  {Stern}, {Benford}, {Blain}, {Carey}, {Cutri}, {Eisenhardt}, {Lonsdale},
  {Mainzer}, {Marsh}, {Padgett}, {Petty}, {Ressler}, {Skrutskie}, {Stanford},
  {Surace}, {Tsai}, {Wheelock}, \& {Yan}}]{jarrett11}
{Jarrett}, T.~H., {Cohen}, M., {Masci}, F.,  {et~al.} 2011, \apj, 735, 112

\bibitem[{{Lang}(2014)}]{lang14}
{Lang}, D. 2014, \aj, 147, 108

\bibitem[{{Lang} {et~al.}(2014){Lang}, {Hogg}, \& {Schlegel}}]{lang14b}
{Lang}, D., {Hogg}, D.~W., \& {Schlegel}, D.~J. 2014, arXiv:1410.7397

\bibitem[{{Levi} {et~al.}(2013){Levi}, {Bebek}, {Beers}, {Blum}, {Cahn},
  {Eisenstein}, {Flaugher}, {Honscheid}, {Kron}, {Lahav}, {McDonald}, {Roe},
  {Schlegel}, \& {representing the DESI collaboration}}]{desi}
{Levi}, M., {Bebek}, C., {Beers}, T.,  {et~al.} 2013, arXiv:1308.0847

\bibitem[{{Mainzer} {et~al.}(2014){Mainzer}, {Bauer}, {Cutri}, {Grav},
  {Masiero}, {Beck}, {Clarkson}, {Conrow}, {Dailey}, {Eisenhardt}, {Fabinsky},
  {Fajardo-Acosta}, {Fowler}, {Gelino}, {Grillmair}, {Heinrichsen}, {Kendall},
  {Kirkpatrick}, {Liu}, {Masci}, {McCallon}, {Nugent}, {Papin}, {Rice},
  {Royer}, {Ryan}, {Sevilla}, {Sonnett}, {Stevenson}, {Thompson}, {Wheelock},
  {Wiemer}, {Wittman}, {Wright}, \& {Yan}}]{neowiser}
{Mainzer}, A., {Bauer}, J., {Cutri}, R.~M.,  {et~al.} 2014, \apj, 792, 30

\bibitem[{{Mainzer} {et~al.}(2011){Mainzer}, {Bauer}, {Grav}, {Masiero},
  {Cutri}, {Dailey}, {Eisenhardt}, {McMillan}, {Wright}, {Walker}, {Jedicke},
  {Spahr}, {Tholen}, {Alles}, {Beck}, {Brandenburg}, {Conrow}, {Evans},
  {Fowler}, {Jarrett}, {Marsh}, {Masci}, {McCallon}, {Wheelock}, {Wittman},
  {Wyatt}, {DeBaun}, {Elliott}, {Elsbury}, {Gautier}, {Gomillion}, {Leisawitz},
  {Maleszewski}, {Micheli}, \& {Wilkins}}]{neowise}
{Mainzer}, A., {Bauer}, J., {Grav}, T.,  {et~al.} 2011, \apj, 731, 53

\bibitem[{{Meisner} \& {Finkbeiner}(2014)}]{meisner14}
{Meisner}, A.~M., \& {Finkbeiner}, D.~P. 2014, \apj, 781, 5

\bibitem[{{Myers} {et~al.}(2015){Myers}, {Palanque-Delabrouille}, {Prakash},
  {P{\^a}ris}, {Yeche}, {Dawson}, {Bovy}, {Lang}, {Schlegel}, {Newman},
  {Petitjean}, {Kneib}, {Laurent}, {Percival}, {Ross}, {Seo}, {Tinker},
  {Armengaud}, {Brownstein}, {Burtin}, {Cai}, {Comparat}, {Kasliwal},
  {Kulkarni}, {Laher}, {Levitan}, {McBride}, {McGreer}, {Miller}, {Nugent},
  {Ofek}, {Rossi}, {Ruan}, {Schneider}, {Sesar}, {Streblyanska}, \&
  {Surace}}]{eboss_qso}
{Myers}, A.~D., {Palanque-Delabrouille}, N., {Prakash}, A.,  {et~al.} 2015,
  \apjs, 221, 27

\bibitem[{{Prakash} {et~al.}(2015){Prakash}, {Licquia}, {Newman}, {Ross},
  {Myers}, {Dawson}, {Kneib}, {Percival}, {Bautista}, {Comparat}, {Tinker},
  {Schlegel}, {Tojeiro}, {Ho}, {Lang}, {Rao}, {McBride}, {Ben Zhu},
  {Brownstein}, {Bailey}, {Bolton}, {Delubac}, {Mariappan}, {Blanton}, {Reid},
  {Schneider}, {Seo}, {Carnero Rosell}, \& {Prada}}]{eboss_lrg}
{Prakash}, A., {Licquia}, T.~C., {Newman}, J.~A.,  {et~al.} 2015,
  arXiv:1508.04478

\bibitem[{{Schlegel} {et~al.}(2015){Schlegel}, {Blum}, {Castander}, {Dey},
  {Finkbeiner}, {Foucaud}, {Honscheid}, {James}, {Lang}, {Levi}, {Moustakas},
  {Myers}, {Newman}, {Nord}, {Nugent}, {Patej}, {Reil}, {Rudnick}, {Rykoff},
  {Ford Schlafly}, {Stark}, {Valdes}, {Walker}, {Weaver}, \& {DECam Legacy
  Survey Collaboration}}]{decals}
{Schlegel}, D.~J., {Blum}, R.~D., {Castander}, F.~J.,  {et~al.} 2015, in
  American Astronomical Society Meeting Abstracts, Vol. 225, American
  Astronomical Society Meeting Abstracts, 336.07

\bibitem[{{Wright} {et~al.}(2010){Wright}, {Eisenhardt}, {Mainzer}, {Ressler},
  {Cutri}, {Jarrett}, {Kirkpatrick}, {Padgett}, {McMillan}, {Skrutskie},
  {Stanford}, {Cohen}, {Walker}, {Mather}, {Leisawitz}, {Gautier}, {McLean},
  {Benford}, {Lonsdale}, {Blain}, {Mendez}, {Irace}, {Duval}, {Liu}, {Royer},
  {Heinrichsen}, {Howard}, {Shannon}, {Kendall}, {Walsh}, {Larsen}, {Cardon},
  {Schick}, {Schwalm}, {Abid}, {Fabinsky}, {Naes}, \& {Tsai}}]{wright10}
{Wright}, E.~L., {Eisenhardt}, P.~R.~M., {Mainzer}, A.~K.,  {et~al.} 2010, \aj,
  140, 1868

\end{thebibliography}

\end{document}